\documentclass[conference,compsoc]{IEEEtran}
\IEEEoverridecommandlockouts

\ifCLASSOPTIONcompsoc

  \usepackage{cite}
\else

  \usepackage{cite}
\fi

\ifCLASSINFOpdf
  \usepackage[pdftex]{graphicx}

\else

  \usepackage[dvips]{graphicx}

\fi

\usepackage{array}

\usepackage{url}

\usepackage{tabularx}
\usepackage{booktabs}

\graphicspath{{./}}

\newcommand{\para}[1]{\vspace{0.3em}\noindent\textbf{#1.}}
\newcommand{\myitem}{\vspace{0.2em}\noindent\textbullet\ }

\begin{document}

\title{CAPED: Context-Aware Privacy Exposure Defense for Mobile GUI Agents\thanks{Artifact availability: We plan to release the CAPED artifact in a future public version.}}

\author{\IEEEauthorblockN{Siyu Shen}
\IEEEauthorblockA{The Chinese University of Hong Kong\\
Email: ss019@ie.cuhk.edu.hk}
\and
\IEEEauthorblockN{Fenghao Xu}
\IEEEauthorblockA{Southeast University\\
Email: fhxu@seu.edu.cn}
\and
\IEEEauthorblockN{Wenrui Diao}
\IEEEauthorblockA{Shandong University\\
Email: diaowenrui@link.cuhk.edu.hk}
\and
\IEEEauthorblockN{\makebox[\textwidth][c]{Kehuan Zhang}}
\IEEEauthorblockA{\makebox[\textwidth][c]{The Chinese University of Hong Kong}\\
\makebox[\textwidth][c]{Email: khzhang@ie.cuhk.edu.hk}}}

\maketitle

\begin{abstract}
Screenshot-based mobile GUI agents can operate ordinary smartphone apps through the same visual interface as a human user, but this capability also turns every screen observation into a privacy boundary. During normal task execution, screenshots may expose contacts, messages, photos, files, recommendations, health cues, and other sensitive context that is unrelated to the user's request. We call this problem incidental visual privacy exposure. It is difficult to address with existing defenses: text anonymization misses many visual and inferential cues, while generic privacy masking can remove the evidence and controls that a GUI agent needs to complete the task.

This paper presents CAPED, a context-aware pre-upload exposure control layer for mobile GUI agents. CAPED is designed as a phone-side protection layer: before screenshots are released to a remote multimodal agent, it extracts task requirements, uses screen context as a privacy prior, parses visible UI elements, and selectively exposes only content needed for the current task while masking incidental private content. We evaluate CAPED on AndroidWorld for broad task utility and with a controlled 28-task seeded privacy evaluation used as a measurement instrument for trajectory-level incidental leakage. In this seeded evaluation, Full CAPED reduces success-conditioned weighted seeded leakage from 0.766 under raw screenshots to 0.268 while preserving high task utility. A broader AndroidWorld run shows a remaining prototype-level utility cost, but the results show that task-driven selective exposure can reduce incidental visual leakage before screenshots are released to a remote GUI agent.

\end{abstract}

\IEEEpeerreviewmaketitle

\section{Introduction}
\label{sec:intro}

Mobile GUI agents operate smartphones through the same visual interface a human uses: they observe screenshots, reason over GUI state, and issue actions such as taps, typing, and swipes. This screenshot-based interface makes automation broadly compatible with existing apps~\cite{appagent,mobile-agent,mobile-agent-v2,uitars,guiowl,androidworld,android-agent-arena,chatgpt-agent,claude-computer-use,droidrun,browser-use,manus-browser-operator}. It also creates a sharp privacy boundary. In many practical deployments, screenshots are processed by an off-device multimodal model or external controller, so the model receives a whole-screen observation even when the user's task requires only a narrow piece of information~\cite{p3-guiguard,webpii,agentdam}.

\begin{figure}[t]
  \centering
  \includegraphics[width=\linewidth]{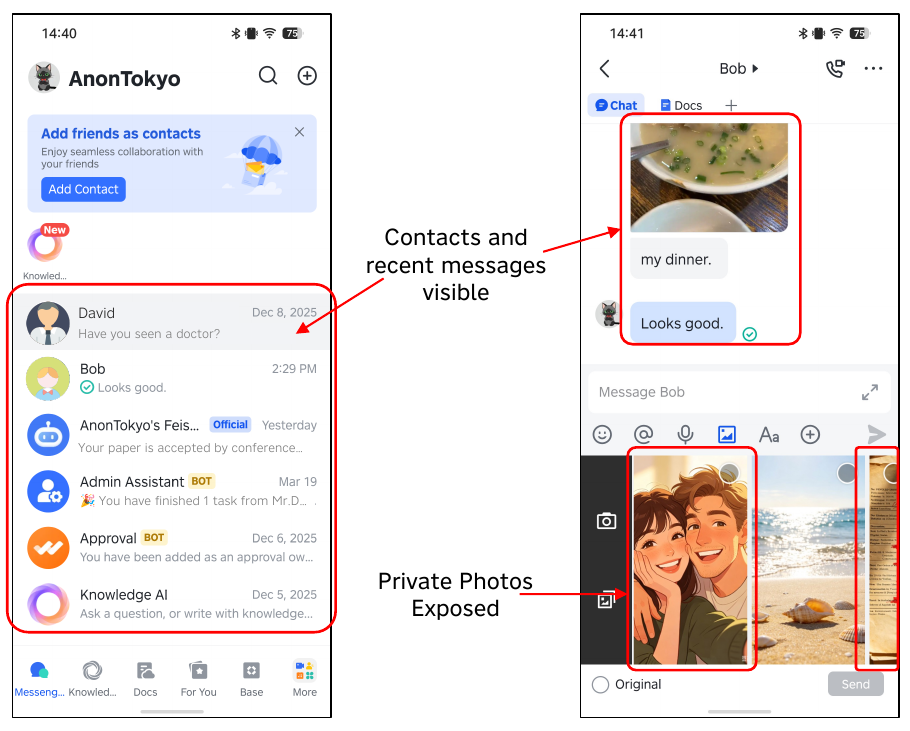}
  \caption{Example of incidental visual privacy exposure in a mobile task flow. Contact previews and private media or file thumbnails can be exposed while the agent searches for task-relevant content.}
  \label{fig:ch3-incidental-example}
\end{figure}

The resulting risk is not limited to explicit personally identifiable information (PII). Consider the task ``send a beach photo to Bob,'' illustrated in Figure~\ref{fig:ch3-incidental-example}. The agent needs to find Bob and the requested photo, but the trajectory may also reveal app choices, recent contacts, message previews, non-target photos, and files visible along the way. These facts are not the target of the task; they are exposed because they happen to share the screen with task-relevant content, echoing broader data-minimization and context-dependent privacy concerns in agent workflows~\cite{agentdam,capid}. We call this \textbf{incidental visual privacy exposure}: task-irrelevant sensitive content that becomes visible to the agent's perception pipeline during normal screenshot-based execution.

Existing defenses do not fully address this setting because the disclosure decision is task-dependent. Text anonymization can replace names, phone numbers, filenames, or message snippets~\cite{p1-anonymization}, but mobile screenshots also reveal faces, avatars, private photos, recommendation patterns, health widgets, and other visual or inferential cues~\cite{vispr,privacyalert,private-attribute-inference,multip2a}. Generic privacy-region masking moves beyond text, but it can destroy the visual evidence a GUI agent needs to navigate, identify targets, and decide where to click; this privacy--utility tension is also central in visual redaction work~\cite{pixel-privacy-utility,visualprivacy-survey}. GUIGuard's online study, for example, reports a sharp task-success drop under full protection and still low success under a task-necessary-aware variant~\cite{p3-guiguard}. These results are not directly comparable to ours, but they expose the same core tension: privacy protection for GUI agents cannot be reduced to detecting sensitive regions and masking them.

This motivates \emph{task-driven selective exposure}. The relevant question is not only whether a screen region is sensitive, but whether that region is needed for this task on this screen. A gallery thumbnail may be essential when the user asks to send that exact photo, but incidental when it appears next to the target. A chat preview may be needed for one conversation while neighboring previews remain unrelated exposure. A useful protection layer must therefore combine the user's task, the screen context, and the visible UI element before deciding what may cross the device--cloud boundary.

\begin{figure*}[t]
  \centering
  \includegraphics[width=0.8\linewidth]{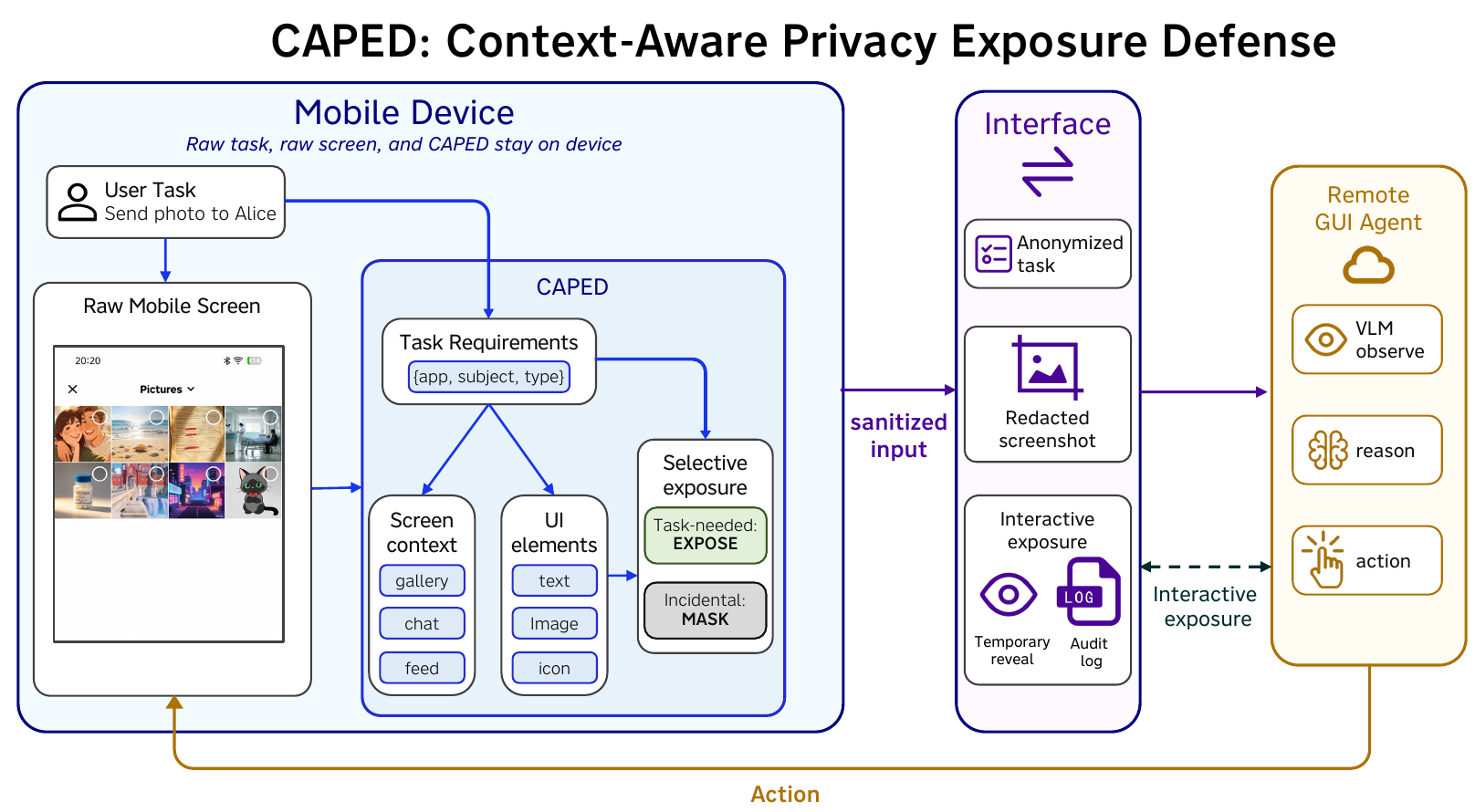}
  \caption{CAPED system overview. In the intended phone-side deployment, raw task text and raw screenshots remain inside the trusted protection boundary. CAPED extracts task requirements, understands the current screen, selectively exposes task-needed content, and sends only a sanitized observation to the remote GUI agent.}
  \label{fig:ch3-caped-overview}
\end{figure*}

We propose \textbf{CAPED} (\textbf{C}ontext-\textbf{A}ware \textbf{P}rivacy \textbf{E}xposure \textbf{D}efense), a phone-side pre-upload protection layer for mobile GUI agents. As shown in Figure~\ref{fig:ch3-caped-overview}, CAPED sits between the raw mobile screen and the remote GUI agent. It keeps the original task, raw screenshots, task understanding, and exposure decisions inside the trusted side of the system, and releases only a sanitized observation: an anonymized task representation and a redacted screenshot. This placement is central to the threat model because the user's instruction itself may contain private names, relationships, locations, or intentions. Our prototype evaluates this pre-upload mediation design inside the AndroidWorld host/server harness; deployment context and runtime costs are reported in Section~\ref{subsec:ch3-runtime-overhead} and Appendix~\ref{app:ch3-deployment-system-details}.

CAPED operationalizes selective exposure through a local decision pipeline. It extracts what the task is allowed to need, uses screen context as a privacy prior, parses visible UI elements, and exposes protected content only when element-level evidence supports task relevance. For a task such as ``Post my cat photos to Xiaohongshu,'' CAPED can expose verified cat photos on a gallery screen while masking unrelated family photos or ambiguous non-target images. The goal is not to censor the interface wholesale, but to preserve the layout, controls, and task-needed evidence while suppressing incidental content carried by individual UI elements.

We evaluate CAPED along two axes. For privacy, we construct a controlled 28-task seeded evaluation spanning mock apps, AndroidWorld-style local apps, and real connected apps, and compare five protection methods using trajectory-level weighted seeded leakage. In this evaluation, No Protection and Text-Only leave most seeded leakage inferable on successful trajectories (WSLR $0.766$ and $0.752$), while Full CAPED reduces success-conditioned WSLR to $0.268$ and maintains high seeded-task utility ($0.929$). For broad utility, we run CAPED in front of GUI-Owl-7B~\cite{guiowl} on AndroidWorld~\cite{androidworld}. On AndroidWorld's 116 ordinary tasks, Full CAPED completes 64 tasks (55.2\%) compared with 77 tasks (66.4\%) for the unprotected GUI-Owl-7B baseline, exposing a measurable prototype utility cost under the current coverage.

\para{Contributions} This paper makes the following contributions:

\myitem We formulate \textbf{incidental visual privacy exposure} in mobile GUI agents: task-irrelevant sensitive content that becomes visible as a byproduct of screenshot-based interaction, including visual and inferential signals beyond textual PII.

\myitem We introduce \textbf{task-driven selective exposure} as a pre-upload privacy principle: expose content only when the current task and screen context justify it, rather than masking or uploading the whole screen.

\myitem We design \textbf{CAPED}, a phone-side exposure-control layer that combines task understanding, screen-context reasoning, UI-element analysis, task-relevance checks, and auditable disclosure before screenshots are released to a remote model.

\myitem We evaluate CAPED on AndroidWorld and a controlled 28-task seeded privacy suite. The results show substantial seeded-leakage reduction under task completion, while also identifying the remaining utility cost and deployment limits of the current prototype.

\section{Problem Setting and Threat Model}
\label{sec:background}

\subsection{Screenshot-Based GUI Agent Workflow}
\label{subsec:ch3-gui-arch}

We model a mobile GUI agent as a perception-action loop. Given a user instruction \(T\), the device captures the current screen \(S_t\), the agent predicts an action \(a_t\), and the device executes that action before capturing the next screen~\cite{appagent,mobile-agent,seeclick,uitars,guiowl}. This interface is powerful because it works across ordinary apps. It also shifts the privacy boundary to the visual observation released at each step, rather than only to the final action.

\subsection{Incidental Visual Privacy}
\label{subsec:ch3-incidental-privacy}

Incidental visual privacy is task-dependent. The same content can be authorized in one task and incidental in another. A target contact, file, photo, message, or health value may be necessary for the user's request, while neighboring contacts, filenames, thumbnails, messages, recommendations, or health cues are not. We therefore distinguish task-required sensitive content, task-irrelevant sensitive content, public or low-risk UI content, and functional controls. CAPED's objective is to preserve the information and controls needed for task execution while minimizing task-irrelevant sensitive exposure.

Incidental exposure includes textual, non-textual visual, and inferential privacy signals. A screenshot can reveal names and message snippets, but it can also reveal faces, personal photos, health widgets, shopping recommendations, locations, or third-party information~\cite{vispr,privacyalert,private-attribute-inference,doxing-lens}. The privacy metric in this paper therefore measures trajectory-level seeded leakage after excluding information that the task itself requires.

\subsection{Threat Model}
\label{subsec:ch3-threat}

The user's device and the local CAPED pipeline are trusted. The remote multimodal model is honest-but-curious: it follows the agent protocol and returns actions, but may retain, inspect, or use uploaded observations beyond the immediate task. We also consider passive exposure through transmitted screenshots or logs. The adversary's goal is to infer private attributes from the sanitized observations released during an agent trajectory. This threat model is consistent with recent evidence that agent observations, memory, and visual inputs can create privacy leakage channels~\cite{eia,agent-memory,mla-trust,private-attribute-inference}.

CAPED is a pre-upload minimization mechanism. It reduces what leaves the trusted phone-side boundary, but it does not control how a provider handles task-needed content after that content has legitimately been exposed. We do not consider compromised operating systems, malicious installed apps, active man-in-the-middle attacks, model compromise, or prompt-injection attacks that manipulate the agent policy.

\section{System Design}
\label{sec:system}

CAPED implements screenshot privacy protection as a local policy-resolution problem. For each visible UI element, it decides whether the element should be exposed to the remote GUI agent given the user's task, the current screen context, and the element modality. This framing is necessary because neither full upload nor full masking is acceptable. Full upload exposes incidental private content that the task does not require~\cite{agentdam,webpii}, while full masking removes visual evidence and controls that a GUI agent needs to act~\cite{p3-guiguard,pixel-privacy-utility}. CAPED therefore does not assign a fixed sensitivity label to an element; it decides whether that element should cross the device--cloud boundary for this task and this screen.

\subsection{Design Goals and Threat Boundary}
\label{subsec:ch3-arch}

\para{Protect before upload} Raw screenshots and the original user task are processed on the phone side of the release boundary before any visual observation is sent to the remote model, following the local-mediation motivation in recent GUI-agent privacy systems~\cite{p1-anonymization,p3-guiguard,webpii,maskclaw}. This rules out designs that ask the cloud agent to decide what should be protected: both the task text and the raw screenshot may contain the private information being protected. Task extraction is therefore part of the trusted local computation, not a preprocessing query to the remote agent. Otherwise, a task such as sending a photo to a named person or checking a private file could disclose the names, relationships, locations, or intentions that CAPED is meant to mediate before any screen policy has been applied.

\para{Preserve task utility} A mobile GUI agent must still see task targets, labels, and navigation controls. CAPED therefore protects at element granularity, preserving usable interface structure while exposing sensitive-looking content when the user's task actually requires it.

\para{Use context as a privacy prior} The same element can require different decisions on different screens. A gallery image is likely personal media, an avatar in a chat list may reveal a social relationship, and a button in a settings page is usually functional. CAPED uses screen context to set the default posture before task requirements unlock specific content, reflecting the broader finding that privacy sensitivity depends on both content and context~\cite{capid,vispr,privacyalert}.

\para{Make masking recoverable} Automatic protection can be conservative or wrong, especially for visually ambiguous targets. CAPED therefore treats exceptional access as an explicit disclosure event rather than silent unmasking: a masked region can be revealed temporarily when an agent action needs it, and the decision is logged.

\subsection{Architecture Overview}
\label{subsec:ch3-architecture-overview}

CAPED is placed between the raw screen and the remote GUI agent, as summarized in Figure~\ref{fig:ch3-caped-overview}. It operates in two phases. The \textit{task analysis phase} runs once per instruction: it de-identifies explicit entities in the task and extracts a structured declaration of what information the agent is allowed to need. The \textit{per-screenshot phase} runs at each agent step: it classifies the screen context, parses UI elements, resolves task-driven exposure decisions, and renders a sanitized screenshot. Across both phases, the remote model receives only the anonymized task representation and the sanitized observation.

The per-screenshot phase combines three local inputs: task requirements, screen context, and parsed UI elements. The policy engine resolves these inputs into an exposure action for each element. Text elements can be exposed through direct subject match or local semantic relevance; image elements require subject-level or navigation-control VQA before protected pixels are released. The redactor then renders exposed original content, masked regions, or anonymized placeholders.

\begin{figure}[t]
  \centering
  \includegraphics[width=0.85\linewidth]{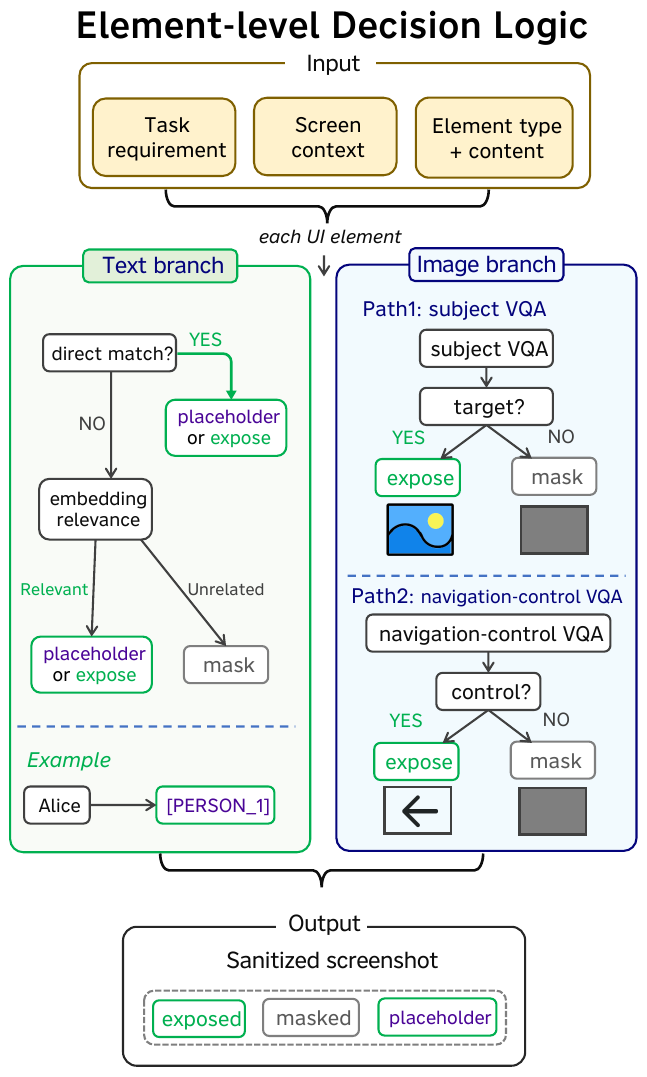}
  \caption{Element-level decision logic in CAPED. For each parsed UI element, CAPED combines task requirements, screen context, and element content, then dispatches text and image elements to modality-specific checks before rendering the sanitized screenshot.}
  \label{fig:ch3-selective-protection}
\end{figure}

\subsection{Local Task Requirements}
\label{subsec:ch3-extractor}

The task requirement extractor declares what information the agent is allowed to need; it does not decide which screen regions are private. This declaration gives the later policy engine a task-specific basis for exposing content. Without it, CAPED would have to choose between exposing all sensitive content for utility or masking it all for privacy. The extractor is deliberately local because the instruction itself may contain private names, locations, intentions, or relationships. The system must know this authorized task scope before it can decide which screenshot regions are safe to release.

\para{Requirement format}
\label{subsubsec:ch3-format}
CAPED represents task needs as a list of \texttt{\{app, subject, type\}} triples. The \texttt{app} field identifies the relevant app by display name. The \texttt{subject} field names the content the agent needs. The \texttt{type} field specifies the kind of subject: \texttt{person}, \texttt{item}, \texttt{date}, \texttt{open}, or \texttt{none}. The key convention is that the triple describes \emph{what} the task needs, not \emph{where} that content will appear on the screen.

The \texttt{open} type denotes an open-ended in-app requirement. It is used when the task target is not a pre-existing named entity that CAPED can match directly, such as creating new content, checking the latest order, or browsing a recent album or sender. This type is necessary because many useful mobile tasks are underspecified until the agent reaches the app surface. It is also the broadest permission in the current schema, so CAPED treats \texttt{open} as scoped exposure within the matched foreground app and current context policy, not as a global permission to reveal the screen. In contrast, specific types such as \texttt{person} or \texttt{item} require name matching, semantic relevance, or visual verification before exposure.

The schema is intentionally compact rather than a complete privacy ontology or app workflow language. A richer schema could encode page scopes, temporal constraints, or app-specific workflows, but requiring the local extractor to predict those details would increase latency and create new failure modes before any screen evidence is available. CAPED instead keeps task extraction coarse and auditable, then defers screen-specific reasoning to runtime context policy and element-level evidence. The extractor itself uses a lightweight local language model with a structured prompt and passes only the parsed JSON requirements downstream. Explicit entities in the task are replaced with typed placeholders. CAPED reuses the same placeholder namespace during screenshot redaction so the remote agent can follow references without receiving the original entity string.

At runtime, CAPED matches each requirement against the foreground app before applying task unlocks. The primary match maps Android package names to app display names and aliases; a token-based fallback handles package variants for common apps. This keeps the task extractor independent of Android package names and page internals. It also prevents a requirement for one app from unlocking similar-looking content in another app merely because the subject text appears on screen.

\subsection{Screen Context and UI Elements}
\label{subsec:ch3-taxonomy}
\label{subsec:ch3-classifier}
\label{subsec:ch3-uiparser}

Screen context provides CAPED's first privacy prior. Mobile apps organize sensitive information by surface: galleries concentrate personal media, chat lists expose social relationships and third-party text, feeds reveal behavioral preferences, and settings or tool screens are mostly functional. Prior visual privacy work similarly shows that private attributes can be tied to visual content and surrounding context rather than only explicit identifiers~\cite{vispr,privacyalert,private-attribute-inference,doxing-lens}. CAPED therefore maps the current screen to a coarse privacy posture before task requirements unlock matched content when exposure is necessary. These policy levels are not the complete exposure categories from Section~\ref{subsec:ch3-incidental-privacy}; they set the default posture before the task-aware resolver separates task-required sensitive content, incidental sensitive content, low-risk context, and functional controls.

\begin{table*}[t]
\centering
\caption{Context policy levels used by CAPED. The levels define default exposure before task-specific unlocking and modality-specific verification.}
\label{tab:ch3-policy-levels}
\footnotesize
\setlength{\tabcolsep}{4pt}
\renewcommand{\arraystretch}{1.12}
\begin{tabularx}{\textwidth}{
  >{\raggedright\arraybackslash}p{1.9cm}
  X
  X
  X}
\hline
\textbf{Policy Level} & \textbf{Default Exposure} & \textbf{Typical Contexts} & \textbf{Task Unlock} \\
\hline
\textsc{Public} & Expose text and images & Settings, tools, games, public content & No unlock needed; this is a low-risk prior, not a guarantee that no private signal exists \\
\textsc{Privacy} & Expose text, protect images & Gallery, maps, files, financial views & Expose matched visual content after verification \\
\textsc{Text-Privacy} & Protect text and images & Messaging, contacts, feeds, profiles, health dashboards, launcher views & Expose matched text or visual content after verification \\
\hline
\end{tabularx}
\end{table*}

The context classifier uses local Android signals whenever possible. Package-level app identity is the primary signal; foreground Activity names and small subpage rules refine mixed-surface apps. The classifier outputs only the coarse policy level in Table~\ref{tab:ch3-policy-levels}. In the current prototype, uncovered contexts fall back to \textsc{Public}; this favors utility, but it is not a privacy-conservative guarantee for unseen private surfaces.

Selective exposure also requires element granularity. CAPED should not hide an entire gallery screen because some thumbnails are private, nor expose a whole chat list because one contact is task-relevant. It therefore parses the screenshot into bounding boxes with coarse types: \texttt{text}, \texttt{icon}, or \texttt{image}. Text can be anonymized or relevance-checked; icons often serve as controls; images in protected contexts often carry private visual content and require verification before exposure. The implementation uses OCR for text regions and a GUI element detector for non-text components~\cite{li2022ppocrv3,uied}, then maps parser outputs to these coarse privacy element types. The parser does not infer privacy by itself; it supplies the element units over which the policy resolver operates.

\subsection{Task-Aware Exposure Resolution}
\label{subsec:ch3-matcher}

Policy resolution is the core of CAPED. For each parsed element, CAPED combines four signals: the context-derived privacy level, the current task requirement, the element modality, and the element content. The output is an exposure decision for this task and this screen: expose the original content, keep the element masked, or render an anonymized placeholder for textual entities. As a result, the same gallery screen can yield different sanitized observations depending on whether the task asks for a specific photo, an open-ended album operation, or no gallery content at all.

We summarize the decision as:
\begin{IEEEeqnarray}{rCl}
\mathrm{exposure}(e \mid s,\tau)
&=& \mathrm{resolve}\bigl(C(s), T(\tau), M(e),\nonumber\\
&& \hphantom{\mathrm{resolve}\bigl(}V(e,\tau)\bigr),
\end{IEEEeqnarray}
where \(s\) denotes the screen and \(\tau\) denotes the task. \(C\) supplies the context prior, \(T\) specifies the task need, \(M\) gives the element modality, and \(V\) provides element evidence from text relevance, subject VQA, or navigation-control verification.

The resolver is implemented as a compact policy table over context level, requirement family, and element modality. It selects one of four actions: expose the element, keep it masked, require subject matching, or run navigation-only verification. Element-specific evidence is used only when the selected action requires matching. This separation keeps the privacy policy deterministic and auditable while still allowing text relevance and VQA to recover task-needed elements that would otherwise remain hidden.

\para{Context default}
\label{subsubsec:ch3-decision}
The screen context first determines the default posture. In \textsc{Public} contexts, elements are exposed because the screen is primarily functional or public. In \textsc{Privacy} contexts, text remains visible but images are masked by default. In \textsc{Text-Privacy} contexts, both text and images are masked by default. Very small images and known safe launcher labels are treated as functional controls so basic navigation remains possible.

\para{Task unlock} A protected element is eligible for exposure only when the foreground app and screen context match a task requirement. Open-ended in-app requirements unlock relevant content within the matched app surface because the task intentionally requires browsing or creating content whose exact target is not known in advance. This is useful for common tasks such as checking the latest order or creating a new note, but it is also a broader privacy scope than a named target. Specific requirements, such as a person, item, or photo subject, require element-level verification before exposure.

\para{Modality-specific verification}
\label{subsubsec:ch3-vqa}
\label{subsubsec:ch3-text-rel}
Text and image elements require different evidence. Text is exposed when the task subject appears directly or when local embedding relevance marks it as task-related. Image elements require VQA: CAPED checks whether a crop contains the requested subject, or whether the crop is a navigation control that should remain usable on protected screens. VQA is used rather than caption or keyword matching because mobile tasks often refer to visually grounded subjects that may not share exact words with a generated caption.

\para{Placeholder post-pass} Even when a text element is exposed for utility, explicit named entities need not be sent verbatim. CAPED replaces detected entities with the same typed placeholders used in the anonymized task, preserving layout and referential structure while avoiding unnecessary PII disclosure.

\subsection{Redaction and Disclosure}
\label{subsec:ch3-protector}
\label{subsec:ch3-disclosure}

A sanitized screenshot must remain a usable GUI observation, not merely a censored image. CAPED's redactor therefore preserves the interface structure seen by the remote agent. Exposed elements keep their original pixels. Masked visual regions are covered with a solid blue-gray overlay, outlined with a border, and marked with an eye icon. Textual entities are replaced with placeholders rather than removed, so the agent can follow references such as \texttt{[PERSON\_1]} across the anonymized task and the redacted screen. After drawing masks, the redactor restores exposed foreground elements so task-relevant labels or controls are not hidden by neighboring protected regions.

Interactive disclosure is a recovery path for conservative masking, not the main privacy mechanism evaluated in this paper. When the agent proposes an action targeting a masked region, CAPED can intercept the action and ask the user whether to reveal that region temporarily. If approved, the original content is revealed only for the requested action and re-masked afterwards; if denied, the action is blocked and the agent must replan. Disclosure events are logged for auditability, but the reported privacy--utility results primarily test automatic selective exposure through task requirements, context policy, element parsing, and verification.

\section{Evaluation}
\label{sec:eval}

CAPED should be judged by a joint criterion: it must preserve task execution while reducing incidental private information exposed to the remote multimodal model. Standard mobile-agent benchmarks measure utility but rarely contain controlled private content. Privacy evaluation, in turn, needs known private facts and a task-required versus incidental distinction. Existing GUI privacy benchmarks target different units: GUIGuard-Bench annotates privacy regions and P-GUI-Evo labels arbitration decisions~\cite{p3-guiguard,maskclaw}, whereas CAPED requires executed sanitized trajectories with seeded facts to measure what task-irrelevant information remains inferable; Appendix~\ref{app:ch3-guiguard-diagnostic} discusses why these benchmarks do not directly apply and how they can still support diagnostics. We therefore use two complementary settings: AndroidWorld measures general task utility, while a controlled seeded privacy evaluation measures trajectory-level incidental leakage under known sensitive content. The evaluation asks three questions:
\begin{itemize}
  \item \textbf{RQ1 (Task utility)}: Can CAPED preserve the agent's ability to complete mobile GUI tasks?
  \item \textbf{RQ2 (Incidental privacy)}: How much task-irrelevant seeded private information remains inferable from the sanitized screenshot trajectory?
  \item \textbf{RQ3 (System tradeoff)}: Which design choices explain the privacy--utility balance, rather than simply obtaining privacy by over-masking the interface?
\end{itemize}

\subsection{Evaluation Suites}
\label{subsec:ch3-eval-suites}

\para{General utility: AndroidWorld} We run CAPED in front of GUI-Owl-7B~\cite{guiowl} on AndroidWorld~\cite{androidworld}, a standardized mobile-agent benchmark with 116 tasks across ordinary Android apps. AndroidWorld is not a privacy benchmark: most tasks do not deliberately expose seeded personal photos, private messages, recommendation feeds, or health panels, and its app contexts are not exhaustively covered by the current CAPED rule base. Its role here is to test whether CAPED breaks ordinary mobile-agent workflows under prototype coverage, not to estimate privacy leakage for unseen private surfaces.

\para{Seeded privacy evaluation} To evaluate privacy directly, we construct 28 privacy-sensitive tasks and execute each task under five protection methods, yielding $28 \times 5 = 140$ task--method runs. The \textbf{mock-app suite} contains nine deterministic tasks across SunShop, MessengerPlus, and HealthTrack. The \textbf{AndroidWorld-style local suite} contains ten tasks in Files, Simple Gallery Pro, and Simple SMS Messenger, using seeded filenames, gallery images, and non-target SMS previews. The \textbf{real connected suite} contains nine tasks in Gmail, Drive, and Photos, adding connected-service email, file, and photo surfaces. Because the connected suite is small, we use it as a realistic trend check rather than as a basis for strong per-app claims.

We use a seeded design because incidental privacy cannot be measured reliably from ordinary task logs alone. The evaluator must know which private facts were intentionally placed in the environment, which of those facts were required by the user task, and which were incidental exposure. Region-level masking counts are also insufficient: the remote model observes a trajectory, so a private fact may be inferred from repeated screens, surrounding metadata, or a combination of visible elements even when no single region is counted as a final leak. The seeded suite is therefore a controlled measurement instrument rather than a community-scale benchmark. It gives auditable privacy measurements for these trajectories, but trades breadth and external validity for task-relevance and leakage ground truth.

\begin{table*}[t]
\centering
\caption{Evaluation suites used in this paper. AndroidWorld measures broad task utility; the seeded privacy evaluation measures incidental leakage under known sensitive content.}
\label{tab:ch3-eval-suites}
\scriptsize
\setlength{\tabcolsep}{4pt}
\renewcommand{\arraystretch}{1.12}
\begin{tabularx}{\textwidth}{
  >{\raggedright\arraybackslash}p{3.0cm}
  >{\centering\arraybackslash}p{1.0cm}
  X
  X}
\hline
\textbf{Suite} & \textbf{Tasks} & \textbf{Privacy Surface} & \textbf{Role in Evaluation} \\
\hline
AndroidWorld &
116 &
Mostly uncontrolled; ordinary phone workflows &
General utility sanity check for No Protection and Full CAPED \\
Mock apps &
9 &
Shopping recommendations, chat lists, contact avatars, health panels &
Controlled privacy evaluation with deterministic seeded leakage \\
AndroidWorld-style local apps &
10 &
Local files, gallery thumbnails/images, SMS previews and contacts &
Seeded privacy evaluation inside reproducible local app workflows \\
Real connected apps &
9 &
Gmail, Drive, and Photos email/file/photo surfaces &
Realistic connected-service trend check for incidental leakage \\
\hline
\end{tabularx}
\end{table*}

\subsection{Compared Methods}
\label{subsec:ch3-eval-methods}

The seeded privacy evaluation compares five methods. \textbf{No Protection} sends raw screenshots to the remote agent. \textbf{Text-Only} replaces textual entities but does not protect visual regions or UI structure. \textbf{CAPED w/o Element Verification} keeps task extraction and context policy but does not selectively rescue masked elements through verification. \textbf{CAPED w/o Task Extraction} applies context protection without task requirements. \textbf{Full CAPED} enables task extraction, context policy, UI parsing, element verification, and redaction rendering. These comparisons separate three claims: whether text anonymization is enough, whether context-only protection collapses into over-masking, and whether task-relevant masked content must be selectively recovered for utility.

\subsection{Metrics}
\label{subsec:ch3-eval-metrics}

\para{Task utility} For AndroidWorld, we report Task Success Rate (TSR), computed by AndroidWorld's built-in task evaluators over 116 tasks. For the seeded privacy evaluation, we report the mean utility score recorded in each run's task-success record. Most seeded tasks are binary, but partial-credit runs with score $0.5$ are retained in the utility mean rather than rounded away.

\para{Trajectory-level seeded leakage} Privacy is measured at the trajectory level because a remote multimodal service observes the full screenshot sequence sent during execution. Counting redacted regions per screenshot would miss cross-screen inference and double-count repeated screens. We therefore use a seeded LLM attacker judge in two stages. It first discovers private facts grounded in visible screenshots, then maps those facts onto a task-specific seeded ontology. The final score is computed deterministically from this fixed ontology mapping, so the model does not choose the severity scale or the denominator of the privacy metric.

For each privacy task, we manually construct the seeded ontology from the private data inserted into the app state. Each ontology item records a seeded fact, whether that fact is required by the task or incidental, and a predefined severity weight. The mock suite covers personalized shopping cues, non-target social identities and messages, and health or medication cues. The local suite covers sensitive filenames, private gallery thumbnails, and non-target SMS previews. The real connected suite covers email, file, and photo surfaces such as financial messages, private correspondence, health images, and legal or medication-related media. We assign weights before judging using a three-level privacy-risk rubric: identity, preference, and generic contextual cues receive weight 1; private message content and pregnancy or parenting cues receive weight 2; and health, medication, mental-health, addiction, financial, legal, and immigration-related facts receive weight 3. Consistent with Section~\ref{subsec:ch3-incidental-privacy}, WSLR excludes task-required seeded information and measures only task-irrelevant seeded sensitive items.

\para{Judge auditability} The attacker judge is used as an evidence mapper, not an unconstrained scorer. It is prompted to report only facts grounded in visible screenshots, avoid speculation, separate task-required information from incidental leakage, and map discovered evidence to the fixed ontology. This design makes the judge conservative in the sense relevant to CAPED: a fact is counted only when the screenshot trajectory provides visible evidence and the mapped item is task-irrelevant under the seeded ontology. We used \texttt{gpt-5.4}, then available as a current OpenAI API model, as the multimodal judge with default decoding~\cite{openai-models}.

To check mapping stability, we repeated the attacker judging three times on a random spot check of 28 existing trajectories covering all three seeded suites and all five methods. The spot check reused the same prompt construction, seeded-ontology validation, and WSLR computation pipeline as the main evaluation. Across 136 incidental seeded item labels, one item changed its leaked/non-leaked mapping across repeats, giving an item flip rate of 0.74\%. The mean trajectory-level WSLR standard deviation was 0.0048. The only nonzero variation came from one HealthTrack trajectory, where one stress-note item was mapped in one repeat but not in the other two. Method ordering within the sampled set was unchanged across all three repeats and under majority-vote aggregation. We also manually inspected the sampled original judge outputs and found their evidence mappings consistent with the visible trajectory evidence.

Let $\mathcal{I}(r)$ denote the set of seeded incidental items applicable to trajectory $r$ after excluding information required by the task. Each item $i \in \mathcal{I}(r)$ has a severity weight $w_i$, and $\lambda_i(r) \in \{0,1\}$ indicates whether the judge marks that item as leaked. We define the \textbf{Weighted Seeded Leakage Rate}:
\[
\mathrm{WSLR}(r) =
\frac{\sum_{i \in \mathcal{I}(r)} w_i \lambda_i(r)}
     {\sum_{i \in \mathcal{I}(r)} w_i}.
\]
Lower WSLR is better. A value of 0 means that none of the task-applicable seeded leakage remains inferable; a value of 1 means that all weighted seeded leakage remains inferable.

\para{Weight sensitivity} The weighted metric reflects the higher privacy cost of health, financial, legal, and similarly sensitive leakage, but the qualitative ordering does not depend on this weighting. A success-conditioned per-run unweighted seeded leakage rate gives the same interpretation, with each successful run contributing equally after excluding task-required seeded items. No Protection and Text-Only remain high at $0.773$ and $0.756$, Full CAPED remains substantially lower at $0.288$, and the two more aggressive ablations are lower still only with worse task utility (CAPED w/o Element Verification $0.140$; CAPED w/o Task Extraction $0.028$).

\para{Success-conditioned privacy} A method should not receive privacy credit merely because it hides the interface so aggressively that the agent fails before reaching private screens. We therefore report privacy averages only over fully successful runs, i.e., runs with recorded utility score exactly 1. Partial-credit runs contribute to utility but are excluded from success-conditioned privacy averages. WSLR and SPS in the main table therefore measure leakage under completed task execution. For readability, we also report \textbf{Seeded Protection Score} $\mathrm{SPS}=1-\mathrm{WSLR}$, where higher is better.

\subsection{Seeded Privacy and Utility Results}
\label{subsec:ch3-seeded-results}

We report the seeded evaluation first because it contains the full five-method privacy--utility comparison under controlled sensitive content.

\begin{figure}[!htbp]
  \centering
  \includegraphics[width=\linewidth]{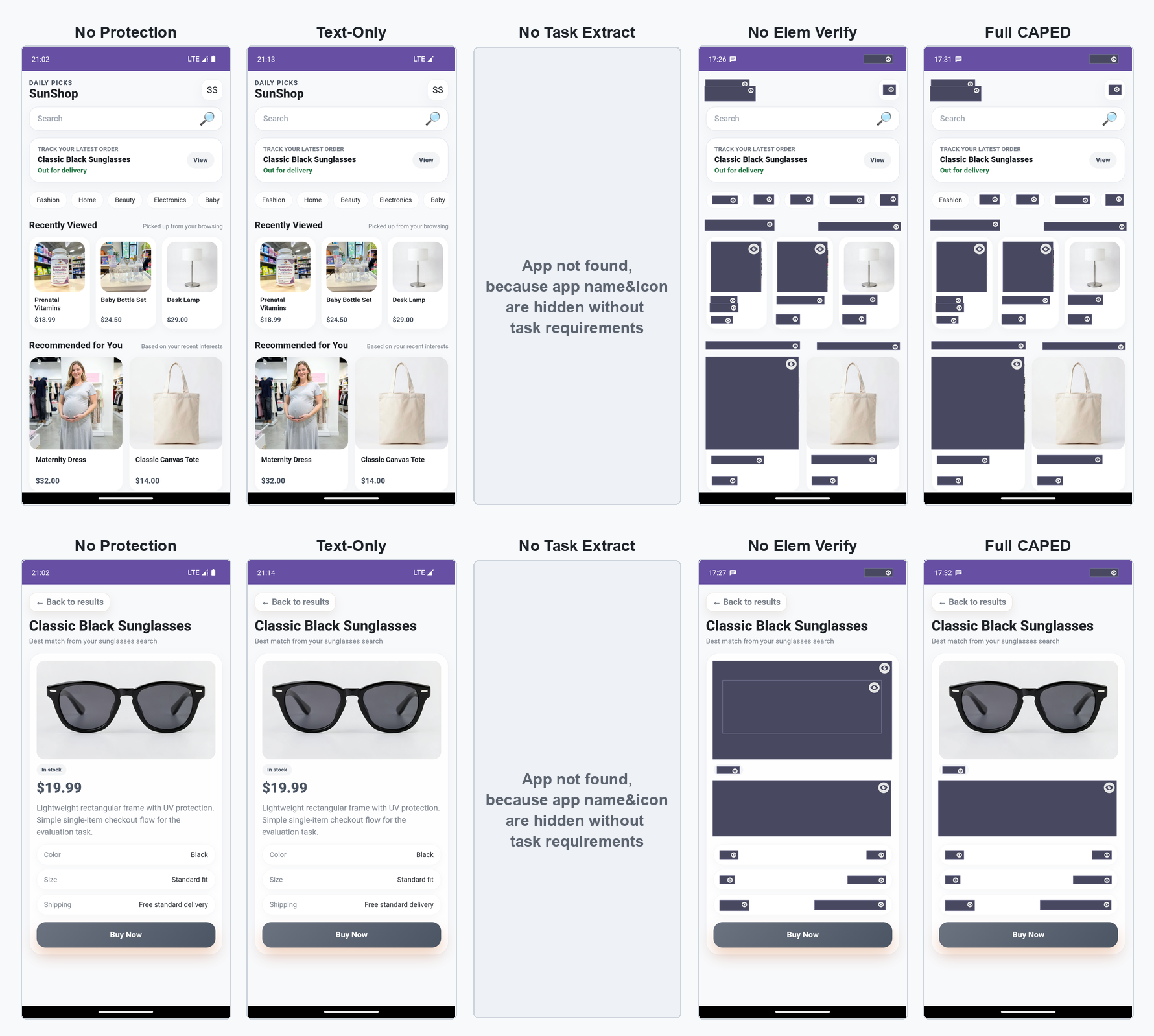}
  \caption{Protection outcome comparison on a SunShop task. The same task is executed under five protection methods; Full CAPED preserves task-relevant product evidence while masking incidental personalized content.}
  \label{fig:ch3-runtime-comparison-sunshop}
\end{figure}

Figure~\ref{fig:ch3-runtime-comparison-sunshop} shows a concrete SunShop run under the five compared methods. No Protection and Text-Only leave personalized recommendation surfaces visible, including pregnancy- and baby-related items that are unrelated to the sunglasses task. Context-only protection without task extraction hides the app entry point and prevents progress. CAPED without element verification protects private regions but can also hide task-relevant product evidence. Full CAPED exposes the task-needed sunglasses while masking incidental recommendations and surrounding private context.

Table~\ref{tab:ch3-seeded-main-results} reports the seeded evaluation results by app setting. Across all 28 seeded tasks, No Protection and Text-Only keep high utility ($0.893$ and $0.911$) but leave most seeded leakage inferable on successful trajectories (WSLR $0.766$ and $0.752$). Full CAPED maintains high utility ($0.929$) while reducing success-conditioned WSLR to $0.268$, corresponding to an SPS of $0.732$. The two ablations clarify the tradeoff: removing element verification reduces leakage further but lowers utility to $0.786$, while removing task extraction yields the lowest leakage only by collapsing utility to $0.321$. In the real connected suite, Full CAPED still reduces WSLR relative to Text-Only ($0.347$ versus $0.402$, a 13.7\% relative reduction), although residual leakage remains substantial.

\begin{table*}[t]
\centering
\caption{Seeded evaluation utility and privacy by app setting. WSLR and SPS are success-conditioned and averaged only over fully successful runs; utility is averaged over all task scores, including partial credit.}
\label{tab:ch3-seeded-main-results}
\scriptsize
\setlength{\tabcolsep}{3pt}
\renewcommand{\arraystretch}{1.05}
\begin{tabularx}{0.88\textwidth}{
  >{\raggedright\arraybackslash}p{3.75cm}
  >{\raggedright\arraybackslash}X
  >{\centering\arraybackslash}p{1.15cm}
  >{\centering\arraybackslash}p{1.1cm}
  >{\centering\arraybackslash}p{1.0cm}
  >{\centering\arraybackslash}p{0.9cm}}
\hline
\textbf{App Setting} & \textbf{Method} & \textbf{Success} & \textbf{Utility $\uparrow$} & \textbf{WSLR $\downarrow$} & \textbf{SPS $\uparrow$} \\
\hline
Mock apps & No Protection & 9/9 & 1.000 & 1.000 & 0.000 \\
Mock apps & Text-Only & 9/9 & 1.000 & 1.000 & 0.000 \\
Mock apps & CAPED w/o Element Verification & 8/9 & 0.889 & 0.204 & 0.796 \\
Mock apps & CAPED w/o Task Extraction & 0/9 & 0.000 & N/A & N/A \\
Mock apps & \textbf{Full CAPED} & \textbf{8/9} & \textbf{0.944} & \textbf{0.217} & \textbf{0.783} \\
\hline
AndroidWorld-style local apps & No Protection & 7/10 & 0.750 & 0.847 & 0.153 \\
AndroidWorld-style local apps & Text-Only & 7/10 & 0.800 & 0.832 & 0.168 \\
AndroidWorld-style local apps & CAPED w/o Element Verification & 6/10 & 0.600 & 0.042 & 0.958 \\
AndroidWorld-style local apps & CAPED w/o Task Extraction & 5/10 & 0.500 & 0.050 & 0.950 \\
AndroidWorld-style local apps & \textbf{Full CAPED} & \textbf{10/10} & \textbf{1.000} & \textbf{0.254} & \textbf{0.746} \\
\hline
Real connected apps & No Protection & 8/9 & 0.944 & 0.433 & 0.567 \\
Real connected apps & Text-Only & 8/9 & 0.944 & 0.402 & 0.598 \\
Real connected apps & CAPED w/o Element Verification & 7/9 & 0.889 & 0.102 & 0.898 \\
Real connected apps & CAPED w/o Task Extraction & 4/9 & 0.444 & 0.000 & 1.000 \\
Real connected apps & \textbf{Full CAPED} & \textbf{7/9} & \textbf{0.833} & \textbf{0.347} & \textbf{0.653} \\
\hline
\end{tabularx}
\end{table*}

Table~\ref{tab:ch3-privacy-utility-tradeoff} summarizes the combined privacy--utility tradeoff over all 28 seeded privacy tasks. Utility is the mean task score across the seeded suite, while SPS is computed from success-conditioned WSLR. No Protection and Text-Only preserve task execution but leave most seeded leakage exposed. The two ablations can appear more private only with lower task utility. Full CAPED gives the best balanced tradeoff in this controlled view.

\begin{table*}[t]
\centering
\caption{Overall privacy--utility tradeoff across the 28-task seeded privacy evaluation. WSLR and SPS are success-conditioned; utility is averaged over all task scores.}
\label{tab:ch3-privacy-utility-tradeoff}
\scriptsize
\setlength{\tabcolsep}{4pt}
\renewcommand{\arraystretch}{1.08}
\begin{tabularx}{0.95\textwidth}{
  >{\raggedright\arraybackslash}p{5.0cm}
  >{\centering\arraybackslash}p{1.15cm}
  >{\centering\arraybackslash}p{1.0cm}
  >{\centering\arraybackslash}p{0.95cm}
  >{\centering\arraybackslash}p{0.9cm}
  >{\raggedright\arraybackslash}X}
\hline
\textbf{Method} & \textbf{Success} & \textbf{Utility $\uparrow$} & \textbf{WSLR $\downarrow$} & \textbf{SPS $\uparrow$} & \textbf{Interpretation} \\
\hline
No Protection & 24/28 & 0.893 & 0.766 & 0.234 & High utility, high leakage \\
Text-Only & 24/28 & 0.911 & 0.752 & 0.248 & Text masking alone is insufficient \\
CAPED w/o Element Verification & 21/28 & 0.786 & 0.124 & 0.876 & Lower leakage, but reduced utility \\
CAPED w/o Task Extraction & 9/28 & 0.321 & 0.028 & 0.972 & Privacy mainly by over-masking and failure \\
\textbf{Full CAPED} & \textbf{25/28} & \textbf{0.929} & \textbf{0.268} & \textbf{0.732} & \textbf{Best balanced tradeoff} \\
\hline
\end{tabularx}
\end{table*}

\para{Takeaways} Three results explain the tradeoff. First, Text-Only remains close to No Protection in aggregate: WSLR is $0.752$ versus $0.766$, showing that incidental screenshot privacy requires more than textual PII removal. Second, CAPED w/o Task Extraction obtains the strongest apparent SPS ($0.972$), but succeeds on only 9 of 28 tasks with mean utility $0.321$; this is privacy by over-masking, not a useful operating point. Diagnostic inspection supports this interpretation: 10 of its 19 failed or partial runs stop before reaching a task-specific private surface. Third, CAPED w/o Element Verification reduces leakage below Full CAPED (WSLR $0.124$ versus $0.268$), but also lowers utility ($0.786$ versus $0.929$). Task-relevant elements therefore need a controlled route back to the agent after context-level masking. Full CAPED's seeded utility should still be read cautiously: it ties No Protection on 23 of 28 tasks, with small task-level variation in both directions, so the aggregate utility ordering is descriptive for this controlled suite rather than a broad agent-capability claim.

\para{Results by app setting} Full CAPED performs best in the local AndroidWorld-style tasks, completing all 10 tasks with WSLR $0.254$. It also retains high utility in the mock apps ($0.944$) while reducing WSLR to $0.217$. The real connected suite is harder: utility falls to $0.833$ and WSLR rises to $0.347$. Compared with Text-Only, this is a 13.7\% relative WSLR reduction ($0.402$ to $0.347$), but the nine-task suite should be read as trend evidence rather than a strong per-app conclusion. These connected tasks expose CAPED to richer Gmail, Drive, and Photos surfaces where task handles, metadata, message previews, thumbnails, and non-target media are tightly interleaved. The utility gap is concentrated in one Photos receipt-album task where CAPED masked thumbnails needed to identify the target image, pointing to a boundary case for target-image verification or interactive disclosure rather than a general collapse of connected-app usability.

\subsection{Residual Leakage}
\label{subsec:ch3-residual-leakage}

Full CAPED does not eliminate all seeded leakage. The remaining errors are concentrated in interpretable boundary cases rather than spread uniformly across tasks. CAPED works best when the user's goal can be represented as a named or otherwise specific target. Open-ended tasks, such as checking the latest item, finding a recent receipt, browsing a recent photo, or creating content whose exact target is not known in advance, require broader in-app exposure and therefore become the main remaining privacy boundary for the current \texttt{\{app, subject, type\}} requirement schema.

The residual cases also reflect dense surfaces where task-relevant handles and incidental private context are tightly interleaved. In the local app suite, residuals include gallery thumbnails related to mental health and medication, plus sensitive filenames. In the mock suite, residual leakage mainly comes from personalized commerce context, non-target contacts, avatars, and health cues. In the real connected suite, residuals concentrate in non-target Gmail messages and Photos thumbnails with financial, health, legal, and medication content. Appendix~\ref{app:ch3-evaluation-diagnostics} gives the detailed residual attribution.

\subsection{General Utility on AndroidWorld}
\label{subsec:ch3-androidworld-results}

The AndroidWorld 116-task run measures whether CAPED remains usable on ordinary mobile-agent workflows beyond the seeded privacy suite. Table~\ref{tab:ch3-androidworld-results} compares the unprotected GUI-Owl baseline with Full CAPED. No Protection completes 77 of 116 tasks (66.4\%), while Full CAPED completes 64 of 116 tasks (55.2\%), an 11.2 percentage-point gap. Pairwise comparison shows 19 regressions where No Protection succeeds but Full CAPED fails, offset by 6 cases where Full CAPED succeeds but No Protection fails.

\begin{table}[t]
\centering
\caption{General utility on AndroidWorld 116 tasks. NaN trials and partial-credit trials are counted as failures, matching the recorded result summary.}
\label{tab:ch3-androidworld-results}
\footnotesize
\setlength{\tabcolsep}{4pt}
\renewcommand{\arraystretch}{1.08}
\begin{tabularx}{0.82\linewidth}{X c c}
\hline
\textbf{Method} & \textbf{Completed Tasks} & \textbf{TSR} \\
\hline
No Protection (GUI-Owl-7B) & 77 / 116 & 66.4\% \\
Full CAPED + GUI-Owl-7B & 64 / 116 & 55.2\% \\
\hline
\end{tabularx}
\end{table}

Manual inspection and targeted diagnostic reruns give a more specific interpretation of these 19 regressions, but we do not use them to adjust the official AndroidWorld TSR. Six were dominated by environment, harness, or app-state failures; six were recoverable under diagnostic conditions but failed in the recorded run or step budget; and seven are best treated as CAPED-affected trajectory cases. Across the inspected cases, direct masking of a task-required control or label was not the dominant failure mechanism. More often, CAPED changed the observation enough to alter the agent's trajectory, consumed additional steps before recovery, or interacted with already fragile app state and timing. The AndroidWorld gap is therefore a real prototype-level utility cost. The diagnostic evidence points mainly to live-loop trajectory sensitivity, step-budget effects, and CAPED-induced observation shifts rather than a simple pattern of required UI being directly occluded.

We do not report AndroidWorld 116-task ablations because the available runs are not controlled enough for a clean method comparison. The seeded evaluation already contains the full five-method ablation study under controlled privacy exposure; AndroidWorld is used here only as the broader utility check for No Protection and Full CAPED.

\subsection{Runtime Overhead}
\label{subsec:ch3-runtime-overhead}

CAPED's protection pipeline runs during agent execution, so its cost matters for practical mobile use. We instrument the Full CAPED AndroidWorld run with per-step timing logs. Across 1,394 measured steps, the complete step takes 10.835 seconds on average in our evaluation setup. Original agent-side components---observation fetch, GUI-Owl inference, and action execution/settling---account for 7.970 seconds. CAPED's online privacy components account for 2.325 seconds, or 22.6\% of measured per-step time.

\begin{figure*}[t]
  \centering
  \includegraphics[width=0.75\textwidth]{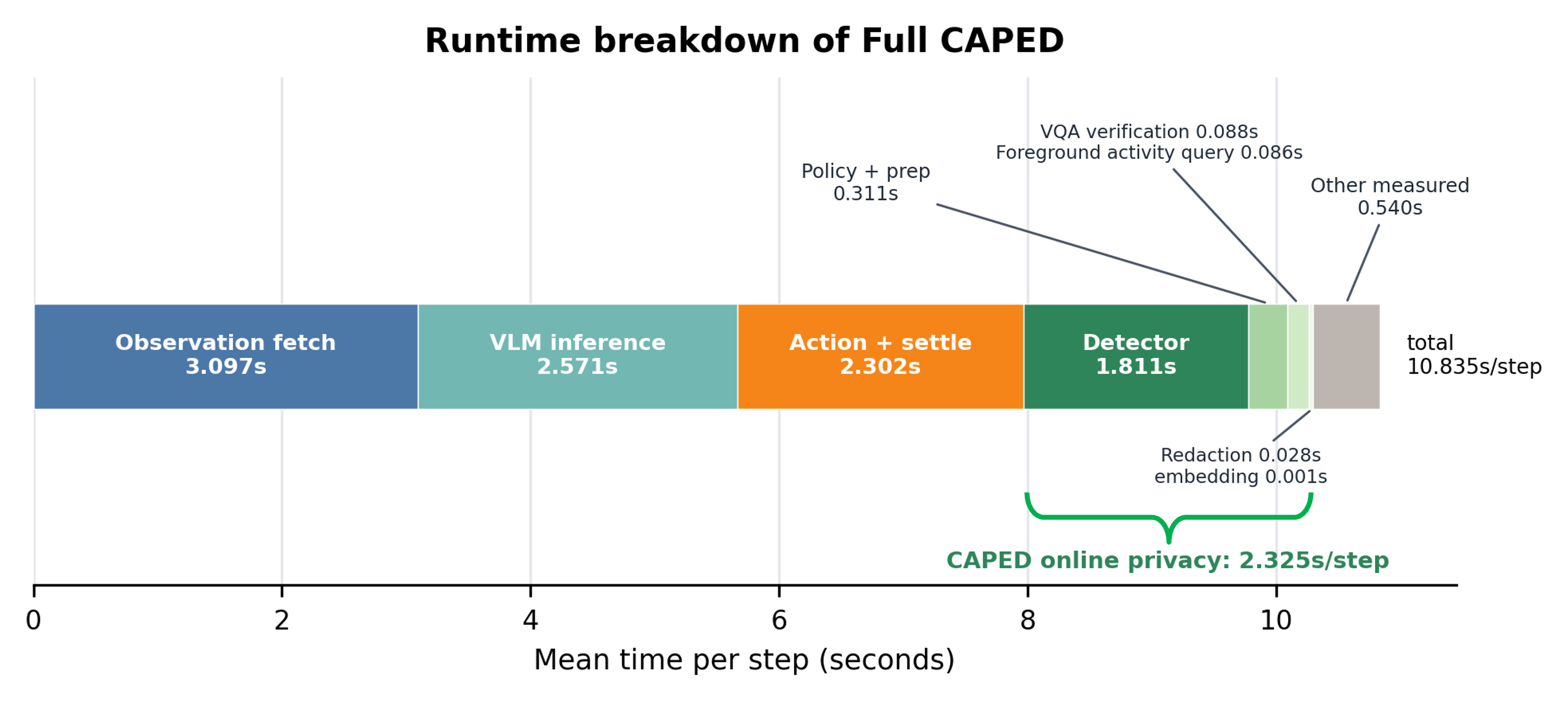}
  \caption{Runtime breakdown of Full CAPED in the live AndroidWorld agent loop. Values are weighted per-step averages over 1,394 measured steps; one-time task extraction is excluded from the per-step bar.}
  \label{fig:ch3-runtime-overhead-breakdown}
\end{figure*}

The dominant CAPED cost is the privacy detector, not the task-aware decision logic or rendering. Detector time averages 1.811 seconds per step, while VQA verification averages 0.088 seconds, embedding relevance is effectively negligible at 0.001 seconds, and redaction rendering averages 0.028 seconds. The one-time task-analysis overhead is separated from per-step latency: requirement extraction averages 5.716 seconds per task over 115 runnable tasks. These measurements show that CAPED is evaluated in the live agent loop rather than only as offline sanitization.

The intended CAPED design is a phone-side protection layer that mediates task interpretation, screen-context reasoning, UI parsing, exposure decisions, and redaction before raw tasks or raw screenshots are released to the remote GUI agent. For engineering convenience, the evaluation prototype runs these CAPED components on the evaluation host and uses an evaluation-side service for heavier model components. The reported measurements should therefore be read as live-loop prototype latency and resource use, not as physical-phone latency or energy. Appendix~\ref{app:ch3-runtime-details} reports additional timing details, and Appendix~\ref{app:ch3-deployment-system-details} gives deployment context.

\section{Related Work}
\label{sec:related}

We situate CAPED against four lines of prior work: mobile GUI agents, agent security and privacy
risks, privacy-preserving GUI-agent execution, and visual privacy or UI understanding. The goal is
to clarify where CAPED fits: it is not a new mobile agent or a generic privacy detector, but a
task-driven release layer for screenshot-based mobile agents.

\subsection{Mobile GUI Agents}
\label{subsec:ch3-related-mobile-agents}

Recent mobile GUI agents demonstrate that smartphone automation through endpoint-level interaction is increasingly practical. Agents can perceive screenshots, reason over interface state, and execute multi-step actions across apps without app-specific APIs. Early systems such as AppAgent~\cite{appagent} and Mobile-Agent~\cite{mobile-agent} frame smartphone operation as human-like interaction with visual interfaces, while SeeClick~\cite{seeclick} and UI-TARS~\cite{uitars} show that screenshot-based GUI grounding and action generation can generalize across mobile, desktop, and web environments. Subsequent systems improve navigation, reflection, long-horizon execution, and deployability through multi-agent collaboration, experience-based self-evolution, and lightweight app-control models~\cite{mobile-agent-v2,mobile-agent-e,appvlm}. Benchmarks and datasets such as AndroidWorld~\cite{androidworld}, AndroidControl~\cite{androidcontrol}, and Android Agent Arena~\cite{android-agent-arena} have also made it possible to measure mobile-agent task completion in more standardized or realistic settings.

The same trend now extends beyond academic prototypes. Computer-use and mobile/browser automation systems have begun to appear in deployed or product-facing forms, including Claude computer use~\cite{claude-computer-use}, Droidrun~\cite{droidrun}, Browser Use~\cite{browser-use}, and Manus Browser Operator~\cite{manus-browser-operator}. Systems such as GUI-Owl~\cite{guiowl} further show that purely screenshot-driven agents can solve a wide range of mobile tasks. This line of work, however, optimizes primarily for task success. It largely assumes that giving the model access to the full screen is acceptable, leaving the privacy cost of that access unaddressed.

\subsection{Mobile Agent Security and Privacy Risks}
\label{subsec:ch3-related-security}

As mobile agents become more capable, recent work has started to analyze the security and privacy risks created by broad screen-level access. Trustworthiness and robustness benchmarks show that multimodal GUI agents can be affected by environmental context, unsafe multi-step behavior, and privacy-sensitive interactive tasks~\cite{envdistraction,mla-trust}. Adversarial studies further show that agent workflows can leak private information through environmental injection, prompt-injection exfiltration, or memory extraction~\cite{eia,prompt-injection-leakage,agent-memory}.

A particularly relevant direction studies privacy awareness: whether an agent can recognize that a requested action or visible content is sensitive and warn the user before proceeding. \emph{Mind the Third Eye!}~\cite{sapabench} introduces SAPA-Bench, a large-scale benchmark for privacy-awareness evaluation in smartphone agents. Its main focus is whether agents can detect privacy-related scenarios, classify their sensitivity, and produce an appropriate warning or confirmation prompt. This is an important complementary problem, but it differs from ours in two key ways: SAPA-Bench evaluates \emph{privacy awareness} and notice-and-consent behavior rather than screenshot sanitization, and it centers on whether an agent should proceed with a sensitive action, not on which on-screen content should be selectively hidden while preserving task execution.

\subsection{Mobile GUI Agent Privacy Protection}
\label{subsec:ch3-related-mobile-privacy}

The closest recent systems to ours study privacy-preserving execution for GUI agents. Several of these systems use or motivate a local--remote hybrid framing: a trusted local or on-device layer mediates screenshots, UI state, or user instructions before a more powerful cloud agent acts~\cite{p1-anonymization,p3-guiguard,webpii,maskclaw}. Their evaluations, however, often instantiate this local layer in emulator or GPU-backed prototype environments rather than as a complete physical-phone deployment. CAPED follows the same architectural trust placement, but makes the prototype placement explicit and keeps the phone-side computation modular: deterministic context policy, UI parsing, a small task extractor, local embedding relevance, conditional crop-level VQA, and redaction.

\para{PII-focused anonymization} Anonymization-Enhanced~\cite{p1-anonymization} proposes a local anonymization layer that replaces sensitive text with deterministic, type-preserving placeholders and remaps actions through a secure proxy. This design is strong for textual PII such as names, phone numbers, and addresses, and it is notable in that task-relevant sensitive values remain usable by the agent without being directly revealed. However, its protection target is still fundamentally \emph{text-centric}. It does not address incidental visual privacy such as recommendation thumbnails, contact avatars, personal photos, or mixed health dashboards in which the privacy risk comes from visible visual content rather than tokenizable strings.

\para{GUIGuard} GUIGuard~\cite{p3-guiguard} is a closely related privacy-preserving GUI-agent system. It detects privacy-sensitive GUI regions, estimates task necessity, and masks protected content before task execution. CAPED differs in its starting point: rather than treating protection mainly as privacy-region detection followed by masking, CAPED formulates the problem as task-driven selective exposure, where task requirements and screen context jointly determine which elements may be released and which constitute incidental visual exposure.

\para{MaskClaw} MaskClaw~\cite{maskclaw} is another recent system that places GUI-agent privacy decisions on the trusted edge before raw screenshots are released. It frames the problem as personalized privacy arbitration, deciding whether an observation should be allowed, masked, or confirmed under user policy and interaction context. CAPED instead treats exposure as a task-necessity problem: it extracts the subjects or scoped app surfaces required by the task, matches visible UI elements against those requirements, and suppresses incidental content that is not needed for the current task.

\subsection{Broader Agent Privacy and Context-Aware Privacy}
\label{subsec:ch3-related-broader}

Beyond mobile agents, several prior works provide useful conceptual context for screenshot privacy, data minimization, context-dependent relevance, and user-mediated disclosure.

\para{Web screenshot privacy} \emph{WebPII}~\cite{webpii} studies visual PII detection in web screenshots for computer-use agents and introduces a fine-grained benchmark together with a real-time redaction model. It strengthens the broader argument that screenshot-based agents need a local privacy layer before cloud inference. However, its focus is still detection-oriented and web-centric. It does not model the mobile GUI setting, trajectory-level incidental exposure, or task-driven decisions about when visually sensitive content should remain accessible.

\para{Agent data minimization and task-scoped disclosure} AgentDAM~\cite{agentdam} studies privacy leakage in autonomous web agents through the lens of data minimization: agents should process private information only when it is necessary for a task-relevant purpose. PrivScope~\cite{privscope} studies a related over-disclosure problem in hybrid local--cloud agents, where a local controller may assemble task-irrelevant working-state context into a cloud-bound payload; it enforces task-scoped disclosure for text payloads rather than screenshot sanitization. These works align with CAPED's selective exposure principle, but they do not address mobile GUI trajectories or visual screenshot sanitization before a remote GUI agent sees the screen.

\para{Context-dependent PII relevance} CAPID~\cite{capid} addresses context-dependent PII detection in question-answering systems, distinguishing between information that is relevant to a query and information that is merely incidental. This closely matches our high-level intuition that privacy relevance is context-dependent. However, CAPID operates entirely in text and does not address visual interfaces, screen context, or agent interaction trajectories.

\para{User-mediated disclosure} PRISM-XR~\cite{prismxr} proposes user confirmation for XR frame uploads to guard against inadvertent capture of private content. Its interactive disclosure model shares conceptual overlap with CAPED's interactive unmasking mechanism. The crucial difference is that PRISM-XR relies on prompting the user whenever sensitive capture occurs, whereas CAPED uses automatic task-aware filtering as the default and reserves user interaction for the narrower case where the agent actively needs a masked region.

\subsection{Vision Models for Privacy and UI Understanding}
\label{subsec:ch3-related-vlm}

Prior work on privacy in general visual content, such as VISPR~\cite{vispr} and PrivacyAlert~\cite{privacyalert}, studies privacy classification in natural photographs. Visual redaction and survey work further emphasizes the privacy--utility tradeoff in hiding sensitive image regions~\cite{pixel-privacy-utility,visualprivacy-survey}, while recent VLM privacy benchmarks show that models can infer or disclose private attributes from visual inputs~\cite{private-attribute-inference,multip2a,doxing-lens}. These systems establish that privacy can be modeled visually rather than only textually, but they are not designed for mobile UI screenshots and do not address the distinctive structure of GUI workflows.

In parallel, UI understanding models such as OmniParser~\cite{omniparser} improve element detection and grounding for agents, but they do not incorporate privacy-aware filtering. CAPED builds on UI parsing as an enabling capability, then adds a task-driven privacy policy layer over mobile UI trajectories.

\section{Discussion}
\label{sec:discussion}

CAPED highlights a practical challenge for pre-upload exposure control in screenshot-based GUI agents: protection requires not merely detecting sensitive content, but deciding whether that content is necessary for the task at hand. A gallery image, chat preview, or health widget cannot be assigned a fixed exposure decision independent of the user's instruction and the screen context in which it appears. The same content category may be task-critical in one trajectory and purely incidental in another.

This paper therefore argues for task-driven selective exposure as a useful abstraction for minimizing incidental visual exposure before mobile GUI-agent screenshots leave the trusted side of the system. Screen context provides a prior about what kinds of private content may be present, while structured task requirements specify what the agent actually needs to see. CAPED combines these two signals before screenshots reach the remote model, aiming to preserve task-relevant evidence while masking incidental content under the evaluated threat model.

This design choice also explains why CAPED performs task interpretation locally. Sending the raw instruction to the remote agent for requirement extraction would leak potentially private names, relationships, locations, or intentions before the protection layer has defined what the task is allowed to reveal. Local extraction adds a one-time startup cost and constrains the expressiveness of the requirement schema, but it preserves the intended mediation point: both task text and screen content are processed before crossing the device--cloud boundary.

\subsection{Limitations and Open Problems}
\label{subsec:ch3-limitations-disc}

\para{Context coverage and unknown apps} CAPED's screen-context policy primarily uses package-level app identity, with Activity and subpage names as limited local hints for mixed-surface apps. This is intentionally a coarse privacy prior, not a robust semantic page classifier. Apps with obfuscated class names, unfamiliar package structures, unsupported workflows, or WebView/custom-view pages may fall through to \texttt{unknown} or may present multiple privacy roles under the same package.

In the current prototype, \texttt{unknown} is mapped to \textsc{Public} as a utility-favoring fallback, because many unknown screens are onboarding, settings, or navigation surfaces. This choice is not privacy-conservative and should be understood as a prototype deployment policy rather than as part of CAPED's privacy guarantee. Our diagnostic found that unknown context was rare in the seeded privacy runs and did not explain non-task-required seeded leakage, so the reported WSLR result is not driven by this fallback. The broader AndroidWorld run, however, contained many unknown-context steps, including ordinary apps whose lists, notes, activities, or task entries can carry private semantics. The current result should therefore be interpreted as protection under covered context rules, not as a guarantee for arbitrary unseen apps.

A stricter deployment could map unknown screens to a protected default or require user confirmation, at the cost of more false masks and more disclosure prompts; quantifying that utility change would require a separate rerun. Broader coverage will require layout-level signals such as uiautomator XML, element-count patterns, lightweight local page classifiers, and real-app seeded benchmarks. When ambiguity remains, the policy should choose the safer context for the user's privacy requirement while making the tradeoff visible in logs.

\para{Element-level ambiguity and dense surfaces} Even when the screen context is correct, element-level decisions can remain ambiguous. Some residual leakage comes from dense surfaces where task-relevant information and incidental private context are tightly interleaved. File lists are a representative case: filenames are functional handles for selecting, moving, or deleting files, but they can also encode medical, financial, immigration, or family information. Health dashboards, shopping pages, messaging lists, gallery grids, and content feeds have similar problems: the target item may be surrounded by recommendations, medication cues, avatars, non-target message previews, or scene-level signals such as a therapy office, legal form, or family setting.

The connected-service results show the same issue in a less controlled form. Gmail, Drive, and Photos surfaces combine task handles with rich non-target messages, metadata, thumbnails, and media grids, so the current prototype reduces but does not robustly suppress incidental leakage there. VQA-based subject verification works best for concrete, visually discriminable targets, and is less reliable for abstract, temporal, or memory-dependent references such as ``the beach photo from last weekend'' or ``my latest receipt.'' CAPED builds its element units from established OCR and GUI-element parsing components, and the residual-leakage attribution points mainly to open-ended task scope and policy exposure rather than parser misses or OCR errors. Future systems may combine finer parsing, stronger non-target suppression, sensitive scene recognition, metadata, temporal ordering, album structure, or user-provided hints, while avoiding over-masking that removes context the agent needs for orientation and navigation.

\para{Requirement-schema expressiveness} CAPED's \texttt{\{app, subject, type\}} requirement schema is deliberately coarse. It does not encode detailed app page scopes, temporal constraints, relational constraints, ordinal references such as ``the second newest file,'' negative constraints such as ``not Alice,'' or separate sensitivity and purpose labels. This simplicity keeps local extraction feasible for small on-device models and avoids requiring the extractor to predict app-internal UI structure, but it limits how precisely CAPED can represent ambiguous multi-step tasks and context-dependent subjects.

Open-ended in-app requirements are the most important boundary case. They are needed when the task target is not named in advance, but our diagnostic shows that they contribute materially to residual leakage. Richer schemas, stricter defaults for open-ended protected surfaces, or user confirmation for broad browsing scopes are plausible future directions. They would also introduce an accuracy--latency--cost tradeoff for the local extractor and would require new experiments rather than a reinterpretation of the current results.

\para{Trajectory-level inference} CAPED makes protection decisions per screenshot, but privacy leakage can accumulate across a trajectory. The same face, username, location, product category, or recommendation pattern may appear in multiple screens, allowing an observer to link identities or infer attributes even if each individual screenshot appears only mildly sensitive. The evaluation already measures leakage at the trajectory level, but the protection policy itself does not yet maintain a trajectory-level privacy state. Extending CAPED with cross-screen privacy budgets, identity linking controls, cumulative exposure accounting, or stricter handling of repeated quasi-identifiers is an important open direction.

\para{Interactive disclosure and user burden} CAPED includes interactive disclosure as a recovery path when automatic protection hides content that the task genuinely needs. The current evaluation, however, does not validate this mechanism as a user-facing workflow. The GUI-Owl-7B agent used in our experiments is a relatively small open-source GUI agent and was not trained or prompted specifically to use CAPED's eye-icon affordance. The current icon may also be too subtle for a general-purpose agent to interpret as an access request.

In the recorded runs, disclosure was triggered only a few times, with automatic approval logs rather than real user decisions, and manual inspection did not find a clear task-beneficial disclosure case. Thus, interactive disclosure should be read as a design mechanism for auditable recovery, not as an evaluated contribution to the reported privacy--utility numbers. Future work should evaluate stronger or disclosure-aware agents, clearer masked-region affordances, and system prompts that explicitly tell the agent that clicking the eye icon requests user-mediated access, while measuring prompt frequency, approval rates, task gains, and user fatigue.

\para{Deployment and runtime validity} CAPED is evaluated as part of a full GUI-agent loop, not as an isolated redaction model. Utility therefore depends not only on whether the correct pixels are exposed, but also on the base agent's robustness to small changes in screenshots, app state, emulator timing, and step budget. The intended deployment is a phone-side pre-upload layer, before raw tasks or raw screenshots are released to the remote GUI agent. For engineering convenience, the prototype evaluation runs CAPED components on the evaluation host and uses an evaluation-side service for heavier model components.

The measured per-step overhead is dominated by detector and UI-parsing cost rather than VQA, embedding, or redaction. Thus, the current artifact evaluates live-loop mediation overhead, but not physical-phone latency, energy, thermal behavior, or mobile runtime integration. A production phone-side system could reduce latency through model distillation, caching stable screen parses, incremental UI updates, mobile NPU acceleration, or platform support for structured UI element access.

\para{Evaluation validity} The current seeded privacy suite is a controlled evaluation under a specific threat model. Its mock apps and AndroidWorld privacy variants make seeded attributes, task relevance, and expected leakage auditable, which is why we use it despite its small scale and custom construction. It does not cover the full diversity of real mobile apps and should not be read as a community-scale benchmark or broad generalization proof. Future evaluation should include more seeded variants for contacts, calendar, finance, maps, gallery-like media selection, and other reproducible privacy contexts, ideally with community benchmarks and blind annotation to improve external validity.

Additional deployment and local task-extraction details appear in Appendix~\ref{app:ch3-deployment-system-details}, and broader impact discussion appears in Appendix~\ref{app:ch3-additional-discussion}.

\section{Conclusion}
\label{chap:conclusion}

This paper studied incidental visual exposure in mobile GUI agents. Screenshot-based agents can operate ordinary apps through the same visual interface as a human user, but that interface also exposes task-irrelevant private content to remote multimodal models. CAPED addresses this tension through task-driven selective exposure: it defines a phone-side pre-upload boundary for task interpretation, screen-context reasoning, and privacy decisions, then sends only sanitized observations to the remote agent.

Our evaluation used AndroidWorld as a broad utility check and a controlled 28-task seeded privacy suite to measure trajectory-level incidental leakage across mock, local, and real connected app settings. CAPED substantially reduced weighted seeded leakage while preserving task utility in the seeded tasks, although the suite is small and the broader AndroidWorld run still showed a prototype-level utility cost. Overall, the results support CAPED's design choice: before screenshots are released to a remote GUI agent, the system should use the current task and screen context to decide which UI elements need to remain visible.

\bibliographystyle{IEEEtran}

\bibliography{references}
\clearpage

\appendices

\section{Deployment and System Details}
\label{app:ch3-deployment-system-details}
\label{app:ch3-problem-details}
\label{app:ch3-system-details}

This appendix collects supplementary details that support CAPED's threat boundary and system design in Sections~\ref{sec:background} and~\ref{sec:system}: the deployment assumption, exposure categories, local task interpretation, and task-requirement representation.

\para{Off-device GUI-agent deployment} CAPED targets a practical architecture in which a local phone-side layer mediates observations before a stronger remote GUI agent acts. AppAgent and Mobile-Agent operate over real-time smartphone screenshots, and Mobile-Agent uses GPT-4V for screenshot-based planning~\cite{appagent,mobile-agent}. Mobile-Agent-v3 centers on GUI-Owl-7B and GUI-Owl-32B~\cite{guiowl}; the released deployment instructions serve these models with server-scale GPU resources~\cite{guiowl7b-huggingface,guiowl32b-huggingface}. Commercial computer-use interfaces similarly define agent loops in which screenshots are returned to a cloud model or external controller~\cite{claude-computer-use}; public reports on Doubao's phone assistant describe a ``user instruction--screenshot--cloud understanding--local execution'' loop for phone control~\cite{doubao-cloud-screenshot}. These examples motivate pre-upload exposure control before observations leave the trusted side of the system.

\para{Exposure categories} The four exposure categories used by CAPED are task-required sensitive content, task-irrelevant sensitive content, public or low-risk UI content, and functional controls. Task-required sensitive content is information the user instruction authorizes the agent to inspect, such as a target contact, file, photo, message, or health value. Task-irrelevant sensitive content is visible during execution but unnecessary for the task; this is the core case of incidental visual privacy exposure. Public or low-risk UI content is content on surfaces that the context policy treats as functional or public by default, although this label is only a prior. Functional controls include buttons, tabs, menus, icons, and layout structure that should remain available when they do not themselves carry private content.

\subsection{Local Task Interpretation}
\label{subsec:ch3-local-remote}

CAPED places the privacy boundary before both screenshot upload and task interpretation. A tempting simplification would be to send the user's instruction to the remote model and ask it to extract task requirements. This would undermine the threat model: the task text may itself contain private information, the remote model is honest-but-curious, and task extraction must happen before the system knows which screen content can be safely exposed. In other words, task understanding is part of the sensitive local computation, not a preprocessing step that can be safely outsourced.

The local task extractor is therefore invoked once per user instruction, rather than once per screenshot. In the current prototype, this component uses Qwen3.5-0.8B with a structured think-first prompt~\cite{qwen3508b-huggingface,qwen35}. The key design point is not the specific model choice, but the placement of task interpretation inside the phone-side release boundary before the remote agent receives any observation.

This placement creates a performance tradeoff. Task extraction is a one-time cost at task start, so it does not add to every screenshot-processing step, but it can still affect perceived latency before the agent begins acting. In our AndroidWorld measurement, the extractor pipeline averaged 5.716 seconds. This cost is acceptable for the current prototype because it protects the original instruction and gives the policy engine a structured basis for selective exposure. Future implementations could reduce it further with platform app-inventory APIs, reusable task templates, constrained parsers for common task forms, smaller local models, or distillation from the current extractor.

\subsection{Task Requirement Examples}
\label{subsec:ch3-task-requirement-examples}

Table~\ref{tab:ch3-req-examples} gives concrete examples of the local task-requirement representation used by CAPED. The main system section describes the policy resolver and redaction behavior; the examples below only illustrate how ordinary user instructions are converted into the compact \texttt{\{app, subject, type\}} schema.

\begin{table}[ht]
\centering
\caption{Example task requirement extractions.}
\label{tab:ch3-req-examples}
\scriptsize
\setlength{\tabcolsep}{4pt}
\renewcommand{\arraystretch}{1.15}
\newcommand{\reqjson}[1]{\begin{minipage}[t]{\linewidth}\ttfamily\scriptsize\raggedright #1\end{minipage}}
\begin{tabularx}{\linewidth}{>{\raggedright\arraybackslash}p{3.15cm} X}
\hline
\textbf{Task} & \textbf{Extractor Output} \\
\hline
``Open the Weather app.'' &
\reqjson{[\{"app":"Weather",\allowbreak{} "subject":"none",\allowbreak{} "type":"none"\}]} \\
``Write a new note in Markor.'' &
\reqjson{[\{"app":"Markor",\allowbreak{} "subject":"open-ended",\allowbreak{} "type":"open"\}]} \\
``Send a message to Bob in WhatsApp.'' &
\reqjson{[\{"app":"WhatsApp",\allowbreak{} "subject":"Bob",\allowbreak{} "type":"person"\}]} \\
\hline
\end{tabularx}
\end{table}

The examples cover three common cases. A navigation-only task produces \texttt{type:none}; an open-ended creation or browsing task produces \texttt{type:open}; and a task with a named target produces a specific subject such as \texttt{type:person}. More complex instructions can produce multiple triples, but CAPED applies the same foreground-app gating and element-level checks described in Section~\ref{sec:system}.

\para{Implementation details} The prototype maps Android package names to app display names and aliases before matching task requirements against the foreground app; a token-based fallback handles common package variants without requiring the local extractor to predict Android package names. OCR regions and GUI component detections are merged into coarse \texttt{text}, \texttt{icon}, and \texttt{image} elements. Large icon-like regions are conservatively reclassified as images so that private thumbnails are not treated as harmless controls, while very small image-like regions are treated as UI chrome for navigation. For task-relevance checks, positive subject VQA requires confidence at least 0.60, local text semantic relevance uses a cosine threshold of 0.60, and launcher app-label matching uses a stricter threshold of 0.65 because labels are short and OCR may be partial.

\section{Seeded Benchmark Construction and Leakage Measurement}
\label{app:ch3-seeded-benchmark-details}
\label{app:ch3-judge}

This appendix details the construction of the seeded privacy suite and the leakage-measurement protocol. Table~\ref{tab:ch3-privacy-task-coverage} reports task coverage rather than results: its purpose is to show the seeded privacy surfaces used for evaluation. The judge protocol then explains how visible evidence is mapped to the fixed seeded ontology used by WSLR.

\begin{table*}[t]
\centering
\caption{Coverage of the 28 seeded privacy tasks. Each task is run under all five compared methods.}
\label{tab:ch3-privacy-task-coverage}
\footnotesize
\setlength{\tabcolsep}{4pt}
\renewcommand{\arraystretch}{1.08}
\begin{tabularx}{\textwidth}{
  >{\raggedright\arraybackslash}p{2.2cm}
  >{\raggedright\arraybackslash}p{2.5cm}
  >{\centering\arraybackslash}p{0.8cm}
  X
  X}
\hline
\textbf{Suite} & \textbf{App} & \textbf{\# Tasks} & \textbf{Task Family} & \textbf{Seeded Incidental Surface} \\
\hline
Mock apps &
SunShop &
3 &
Browse, search, and check order state in a shopping app &
Pregnancy and parenting recommendations, personalized commerce history \\
Mock apps &
MessengerPlus &
3 &
Read, reply, and search in a messaging app &
Non-target contacts, avatars, and sensitive message previews \\
Mock apps &
HealthTrack &
3 &
Inspect steps, sleep, and medication reminders &
Health condition, medication, dosage, schedule, and stress cues \\
\hline
AndroidWorld-style local apps &
Files &
2 &
Move or delete a target file &
Unrelated sensitive filenames about medical, financial, visa, and school-record topics \\
AndroidWorld-style local apps &
Simple Gallery Pro &
4 &
Copy, open, find, or share a target image &
Unrelated medical, medication, therapy, legal-document, and personal thumbnails \\
AndroidWorld-style local apps &
Simple SMS Messenger &
4 &
Reply, forward, search, or compose in local SMS workflows &
Non-target previews and contacts about oncology, therapy, immigration, mortgage, and divorce \\
\hline
Real connected apps &
Gmail &
3 &
Open, search, or read a target email &
Non-target financial, medical, and private-message email context \\
Real connected apps &
Drive &
3 &
Open, search, or share a target cloud file &
Non-target cloud-file names and document surfaces \\
Real connected apps &
Photos &
3 &
Open or share a target album photo &
Non-target medical, legal, medication, and private media thumbnails \\
\hline
\end{tabularx}
\end{table*}

\para{Attacker judge protocol} The attacker judge follows a two-stage protocol. In the discovery stage, the judge observes the full screenshot trajectory and lists private or sensitive facts grounded in visible evidence. In the mapping stage, it maps those facts to a task-specific seeded leakage ontology. The scoring script then computes WSLR deterministically from the mapped items and predefined weights.

\para{Discovery prompt excerpt} The judge is instructed to infer private or sensitive facts from the screenshot trajectory, but only when grounded in visible screenshot evidence. It is explicitly told not to speculate beyond visible content, to prefer concrete facts over broad summaries, and to return structured JSON containing the fact, subject, category, evidence, confidence, and screenshot indices.

\para{Mapping prompt excerpt} The judge receives the discovered facts and the seeded leakage items for the task. It marks a seeded item as leaked only when supported by visible screenshot evidence, excludes task-required information from incidental leakage, and returns structured JSON with leaked status, evidence, screenshot indices, and supporting facts.

\para{Seeded ontology example} For a SunShop task, the ontology includes seeded incidental items such as pregnancy-related products, baby-care products, maternity clothing, and personalized recommendation cues. For each item, the ontology specifies an item id, privacy category, description, subject, whether it is task-required, and an evidence rule. Category weights are predefined before judging; for example, contact identity and shopping preference have lower weights than private message content, pregnancy or parenting cues, medication, health-condition, mental-health, financial, legal, or immigration cues.

\para{Ontology coverage diagnostic} The seeded ontology defines the primary metric denominator, but the judge also records discovered private facts that do not map to a seeded item. We use these out-of-ontology facts as an audit signal rather than as part of WSLR. This is especially useful for connected apps, where real Gmail, Drive, and Photos surfaces may expose recurring private cues beyond the seeded facts. In the current evaluation, such records are used to check whether the fixed ontology misses systematic leakage categories; they do not change the reported seeded scores.

\section{Evaluation Diagnostics}
\label{app:ch3-evaluation-diagnostics}

This appendix collects diagnostics that support the interpretation of the seeded results. The ablation table links each compared method to the system component it removes; the residual and context tables explain where leakage remains and how much the results depend on context rules or disclosure prompts.

\begin{table*}[t]
\centering
\caption{Ablation interpretation by system component. Evidence is drawn from Table~\ref{tab:ch3-seeded-main-results}.}
\label{tab:ch3-ablation-design-link}
\footnotesize
\setlength{\tabcolsep}{4pt}
\renewcommand{\arraystretch}{1.08}
\begin{tabularx}{\textwidth}{
  >{\raggedright\arraybackslash}p{3.0cm}
  >{\raggedright\arraybackslash}p{3.0cm}
  X
  X}
\hline
\textbf{Method} & \textbf{Component Removed or Weakened} & \textbf{Expected Failure Mode} & \textbf{Observed Evidence} \\
\hline
Text-Only &
Screen-context policy and visual element protection are absent &
Text replacement misses visual, structural, and contextual leakage in screenshots &
Utility remains high ($0.911$), but WSLR stays close to No Protection: $0.752$ versus $0.766$ \\
CAPED w/o Task Extraction &
Local task requirements and task-driven unlock are absent &
Context protection can become too conservative, producing apparent privacy by hiding useful content &
Mean utility falls to $0.321$ and only 9 of 28 tasks succeed, even though apparent SPS is high \\
CAPED w/o Element Verification &
Element-level verification and selective rescue are absent &
Protected elements that are actually task-relevant cannot be recovered reliably &
WSLR falls to $0.124$, but utility also falls to $0.786$; Full CAPED trades some leakage for better task completion in this suite \\
Full CAPED &
All core components are enabled &
Task-needed content can be exposed while much seeded incidental leakage is suppressed &
Full CAPED maintains $0.929$ mean utility in this seeded suite and reduces success-conditioned WSLR to $0.268$ \\
\hline
\end{tabularx}
\end{table*}

\para{Seeded utility and success-conditioning checks} Several diagnostics support the interpretation of the seeded results without requiring another table. Full CAPED and No Protection have the same utility on 23 of 28 seeded tasks; Full CAPED scores higher on three tasks and lower on two tasks, for a net one-point task-score difference. This means the seeded utility ordering should be read as descriptive trajectory variation in this controlled suite, not as a broad claim that CAPED improves agent capability.

The success-conditioning checks support the same reading. Full CAPED's all-run WSLR is close to its success-conditioned WSLR ($0.280$ versus $0.268$), and none of its three failed or partial runs failed before reaching a task-specific private surface. By contrast, CAPED without task extraction fails or receives partial credit before reaching such a surface in 10 of 19 failed or partial runs, supporting the interpretation that its low leakage partly comes from over-masking before useful task progress.

\para{Residual attribution} Table~\ref{tab:ch3-appendix-residual-attribution} attributes residual leakage in successful Full CAPED runs. The attribution is based on leaked seeded facts, not raw parser boxes, all redacted regions, or all visible regions.

\begin{table*}[t]
\centering
\caption{Residual leakage attribution in successful Full CAPED runs.}
\label{tab:ch3-appendix-residual-attribution}
\footnotesize
\setlength{\tabcolsep}{4pt}
\renewcommand{\arraystretch}{1.08}
\begin{tabularx}{\textwidth}{
  >{\raggedright\arraybackslash}p{3.4cm}
  >{\centering\arraybackslash}p{1.0cm}
  >{\centering\arraybackslash}p{1.2cm}
  X}
\hline
\textbf{Primary Residual Source} & \textbf{Facts} & \textbf{Weight} & \textbf{Interpretation} \\
\hline
Open-ended in-app scope &
13 &
33 / 80 &
Broad browsing or creation scope exposed surrounding non-target content; among successful Full CAPED trajectories, the broad-scope subset has higher WSLR than the non-broad-scope subset ($0.614$ versus $0.191$) \\
Policy exposure &
11 &
30 / 80 &
A parsed element was exposed by the current task/context policy even though it later mapped to incidental seeded leakage \\
Rendering or perception artifact &
4 &
10 / 80 &
Masking/restoration behavior or saved processed artifacts left protected content visible \\
Other VQA, context, or subject-matching limits &
5 &
7 / 80 &
Ambiguous subject grounding, file/context handling, or manual-judgment uncertainty explains the remaining residuals \\
\hline
\end{tabularx}
\end{table*}

\para{Context and disclosure diagnostics} Table~\ref{tab:ch3-context-disclosure-diagnostics} bounds how much the reported seeded results depend on context-rule details or disclosure prompts.

\begin{table*}[t]
\centering
\caption{Context-prior, unknown-context, and interactive-disclosure diagnostics.}
\label{tab:ch3-context-disclosure-diagnostics}
\footnotesize
\setlength{\tabcolsep}{4pt}
\renewcommand{\arraystretch}{1.08}
\begin{tabularx}{\textwidth}{
  >{\raggedright\arraybackslash}p{3.1cm}
  >{\raggedright\arraybackslash}p{4.3cm}
  X}
\hline
\textbf{Diagnostic} & \textbf{Observed Result} & \textbf{Implication} \\
\hline
Context-prior sensitivity &
210 recorded Full CAPED seeded steps; Activity or subpage rules refined the internal context label in 41 steps, but the coarse \textsc{Public}/\textsc{Privacy}/\textsc{Text-Privacy} level was unchanged in all 210 steps &
The reported seeded results are not driven by fragile Activity-specific changes to the three-level privacy posture \\
Unknown context in seeded runs &
0 of 53 mock steps, 1 of 60 AndroidWorld-style local steps, 0 of 32 expandOther steps, and 3 of 65 expandGoogle steps; no non-task-required seeded leakage was attributed to the unknown fallback &
The seeded WSLR result is mostly measured under covered context rules, so it should not be read as an unseen-app privacy guarantee \\
Unknown context in AndroidWorld utility run &
602 of 1,341 recorded steps (44.9\%) across 65 of 106 tasks with parsed context logs &
This prototype uses a utility-favoring \texttt{unknown}$\rightarrow$\textsc{Public} fallback; stricter unknown defaults would require additional masking and a separate utility rerun \\
Disclosure logs &
Three \texttt{auto\_approve} events: one seeded mock event, one seeded expandOther event, and one AndroidWorld event &
The reported privacy--utility results come from automatic selective exposure rather than systematic use of interactive disclosure \\
\hline
\end{tabularx}
\end{table*}

\section{Runtime Overhead Details}
\label{app:ch3-runtime-details}

\para{Per-step overhead} Figure~\ref{fig:ch3-runtime-overhead-breakdown} in the main text summarizes the measured live-loop overhead. Across 1,394 AndroidWorld steps, the measured step total averages 10.835 seconds. The original GUI-agent loop accounts for 7.970 seconds, while CAPED's online privacy components account for 2.325 seconds. The dominant CAPED-side cost is OCR, UI parsing, and detection at 1.811 seconds per step; VQA verification averages 0.088 seconds, embedding relevance 0.001 seconds, and redaction rendering 0.028 seconds.

\para{Task-start extraction overhead} Task requirement extraction is a one-time task-start cost rather than a per-screenshot cost. Over the 115 AndroidWorld goals that reached the privacy hook extraction boundary, the extractor averaged 5.716 seconds, with median 5.634 seconds and range 5.344--8.485 seconds. This measurement excludes emulator startup, screenshots, NER, and per-step privacy processing.

\para{Prototype environment} The AndroidWorld utility run uses an Android emulator environment consistent with the AndroidWorld setup, with CAPED integrated into the live GUI-agent loop rather than applied as offline post-processing. The evaluation prototype runs orchestration and screen-processing components on the evaluation host and uses an evaluation-side service for heavier model components. The remote agent is GUI-Owl-7B; the local task extractor uses Qwen3.5-0.8B~\cite{qwen3508b-huggingface,qwen35}; text relevance uses EmbeddingGemma~\cite{embed300m-huggingface,embedding_gemma_2025}; and the screen parser uses OCR and UI-element detection components, including PaddleOCR~\cite{li2022ppocrv3} and a UIED-based detector~\cite{uied}. These implementation choices are prototype placements for measuring live-loop mediation overhead, not a claim of final physical-phone latency or energy.

\section{External Benchmark and Broader Context}
\label{app:ch3-external-broader-context}
\label{app:ch3-guiguard-diagnostic}
\label{app:ch3-additional-discussion}

This appendix contains supplementary context that is not part of CAPED's primary evaluation: external benchmark diagnostics and broader impact considerations.

\subsection{External Benchmark Diagnostics}

Existing GUI privacy benchmarks are useful diagnostics but do not directly replace CAPED's seeded trajectory evaluation. GUIGuard-Bench's Android subset~\cite{p3-guiguard} provides oracle privacy-region annotations, including bounding boxes, risk levels, privacy categories, and task-essential labels. CAPED, however, does not make decisions over oracle privacy regions. It parses each screenshot at runtime into OCR text regions, image crops, icons, and controls, then resolves exposure using task requirements and screen context; the annotation granularity is therefore different. GUIGuard-Bench also does not provide the Android package, Activity, or subpage state used by CAPED's context policy and app-specific task gating. Finally, it lacks the post-sanitization agent trajectory needed to measure WSLR.

P-GUI-Evo~\cite{maskclaw} is closer to policy arbitration: it uses reconstructed GUI decision samples with persona, app, instruction or agent-intent context, interaction context, visible sensitive evidence, and reference Allow/Mask/Ask-style outcomes. Its central mismatch with CAPED is the label semantics. P-GUI-Evo evaluates which arbitration action is appropriate under a policy and interaction context, whereas CAPED measures which task-irrelevant private facts remain inferable after task-conditioned screenshot sanitization. A single reconstructed decision sample is also not a multi-step mobile-agent trajectory, and the dataset does not provide CAPED's task-requirement representation, runtime parsed elements, sanitized observations, or downstream agent continuation.

\para{Offline alignment diagnostic} We therefore use GUIGuard-Bench only as an external alignment diagnostic. The Android subset we inspected contains 68 task folders, 1,239 screenshot steps, and 9,032 region labels, each with a bounding box, risk level, and task-essential flag. Our adapter assigns each screenshot a coarse context, applies CAPED's policy table to the annotated regions, and treats no-risk non-essential labels as neutral rather than privacy targets. The resulting counts mix CAPED policy limitations with missing runtime components and benchmark-denominator mismatch, so we do not report them as a primary CAPED privacy score.

\para{Takeaway} The diagnostic still surfaced useful failure modes: broad exposure of semantically sensitive file-list text, content-feed exposure under open-ended browsing or creation tasks, and overly broad rescue of person-name or subject-matched regions when the offline adapter pools task subjects without CAPED's full runtime gating. These observations are consistent with the residual-leakage analysis in Section~\ref{subsec:ch3-residual-leakage}. We therefore use GUIGuard-Bench as public-data support for diagnosis, while the controlled seeded benchmark remains the primary trajectory-level privacy--utility measurement.

\subsection{Broader Impact}
\label{subsec:ch3-impact}

\para{Benefits} CAPED reduces the amount of private visual context exposed to cloud-based GUI agents, gives users a more fine-grained control surface than all-or-nothing screen sharing, and records disclosure events for auditability. These properties are particularly important for mobile devices, where a single screen may contain first-party data, third-party messages, and behavioral traces in close proximity.

\para{Risks} CAPED can also create a false sense of security if users assume that all private information is protected. Context classification may be incomplete, VQA may miss a task-relevant subject, and trajectory-level inference remains only partially addressed. CAPED also does not protect information that the user task genuinely requires exposing, nor does it control how a remote model provider later handles task-needed content that was legitimately sent. Over-masking may reduce agent autonomy or cause users to approve disclosures repeatedly without careful review. Finally, third-party privacy remains a social problem as well as a technical one: a user may delegate a task involving contacts or messages without those third parties' awareness.

\para{Mitigations} CAPED's audit logs, disclosure prompts, and explicit scope limitations are intended to make these risks visible rather than hidden. A production deployment should expose unknown-context decisions, disclosure history, and protection scope clearly to users, and should allow stricter defaults for users or organizations that prefer stronger privacy over smoother automation. Provider-side privacy commitments, retention limits, or trusted execution would be complementary mechanisms for task-needed content after it has legitimately crossed the device boundary; they are outside CAPED's current technical guarantee.

\end{document}